\documentclass[prd,nofootinbib,onecolumn]{revtex4}
\usepackage{amsmath}
\usepackage{amssymb}
\usepackage{graphics}
\usepackage{epsfig}

\newcommand\nn{\nonumber}
\newcommand\ba{\begin{eqnarray}}
\newcommand\ea{\end{eqnarray}}

\begin{document}

\title{Generalized polarizability of neutral pions of the process
$e^- e^+ \to \pi^0\pi^0\gamma$ in NJL model}

\author{A.~I.~Ahmadov }
\email{ahmadov@theor.jinr.ru}
\affiliation{Joint Institute for Nuclear Research, Dubna, Russia}
\affiliation{Institute of Physics, Azerbaijan National Academy of Sciences, Baku, Azerbaijan}

\author{E.~A.~Kuraev}
\email{kuraev@theor.jinr.ru}
\affiliation{Joint Institute for Nuclear Research, Dubna, Russia}

\author{M.~K.~Volkov}
\email{volkov@theor.jinr.ru}
\affiliation{Joint Institute for Nuclear Research, Dubna, Russia}

\date{\today}

\begin{abstract}
Differential distributions in the $\pi_0\pi_0\gamma$ system created in the annihilation channel
of an electron-positron collision are considered.
The energy fractions of the pions (Dalitz-plot)
distribution are presented in a general form and in approximation of intermediate
vector mesons (excited and ordinary ones).It is pointed out that in relevant
experiments the generalized polarizability of the neutral pion can be measured.
Numerical illustrations are presented.
\end{abstract}

\maketitle

\section{Introduction}
\label{sect1}

In a recent paper \cite{pigamm} the process of creation of $\pi^0\omega$ in annihilation of
an electron-positron pair was considered in the framework of the Nambu-Jona-Lazinio model. 
The result obtained turns out to be in very satisfactory agreement with theoretical estimations based
on vector meson dominance and experimental measurement of this process with subsequent
decay of the $\omega$ meson to $\pi_0\gamma$ \cite{Ahmet03}.
In this experiment, the dependence of the total cross section on the total energy of initial
electron and positron in the center of mass reference frame was measured.

It is the motivation of this paper to pay attention to the differential distributions
in momenta of final particles since they carry information on the properties of the neutral
pion such as polarizability.

In this paper, the process of annihilation of $e^- e^+ \to \gamma^* \to \pi^0\pi^0\gamma$ is considered.
It can be obtained from the process $e^- e^+ \to \pi^0\omega$, if $\omega \to \pi^0\gamma$.
Hence, we can use the results of \cite{pigamm}.

Measuring of the process
\begin{gather}
e^+(p_+)+e^-(p_-)\to \gamma(q)\to \pi_0(q_1)+\pi_0(q_2)+\gamma(k)
\end{gather}
provides a possibility to investigate the crossing process to the Compton scattering
on neutral pion in the case when initial photon is of mass shell. Also, the role of
excited vector mesons and the generalized polarizabilities of neutral pion can be
studied. To article, this process can be useful for research Compton effect.

The dependence of the total cross section on the total energy of the initial electron and positron
in the center of mass reference frame first was measured in \cite{Ahmet03}.

This process allows us to describe the process where $e^- e^+ \to \pi^0\pi^0\gamma$,
when $\omega \to \pi^0\gamma.$
\section{Invariant Structure of the process of $e^- e^+ \to \pi^0\pi^0\gamma$}
In paper \cite{EK85}, the matrix element and the
differential cross section was obtained in terms of irreducible tensor structures:
\ba
M=8\pi m\sqrt{4\pi\alpha}\frac{1}{s}J^\mu e^\nu [a_1L^{(1)}_{\mu\nu}+a_2L^{(2)}_{\mu\nu}+
a_3L^{(3)}_{\mu\nu}],
\ea
$a_1, a_2$, and $a_3$ define the properties of polarizability neutral pions,
where $J_\mu=\bar{v}(p_+)\gamma_\mu u(p_-)$ is the electromagnetic current,
\ba
L^{(1)}_{\mu\nu}=(q k)g_{\mu\nu}-k_\mu q_\nu, \nn \\
L^{(2)}_{\mu\nu}=(k q)Q_\mu Q_\nu-(Q q)(Q_\mu q_\nu+Q_\nu k_\mu)+(Q q)^2g_{\mu\nu}; \nn \\
L^{(3)}_{\mu\nu}=(Q q)(q^2g_{\mu\nu}-q_\mu q_\nu)+Q_\nu(k q q_\mu-q^2 k_\mu), \nn \\
Q=\frac{1}{2}(q_1-q_2), \,\,q_1^2=q_2^2=m^2,\,\, k^2=0, \,\,q^2=s=4E^2,
\ea
$m$ is the pion mass, $E$ is the energy of electron (positron), and $2E$ is the total energy in the center
of mass reference frame.
The tensors $L^{(i)}_{\mu\nu}$ obey the gauge conditions
\ba
L^{(1)}_{\mu\nu}q_\mu=0;\,\,\,\,\,L^{(i)}_{\mu\nu}k_\nu=0, \,\,\,i=1,2,3.
\ea
The coefficients $a_i$ are some functions of energy fractions of pions $x_{1,2}=2q_iq/s$. Moreover,
due to Bose statistics they obey the symmetry conditions
\ba
a_1(x_1,x_2)=a_1(x_2,x_1),a_2(x_1,x_2)=a_2(x_2,x_1),a_3(x_1,x_2)=-a_3(x_2,x_1).
\ea
The parameter $a_1$ coincide in static limit $q^2=0$ with neutral pion
electric polarizability $\alpha_1=m^3\alpha_0$, $|\alpha_0|\approx  6.6\times 10^{-43}cm^3$
\cite{MKV86},
The parameters $a_2,a_3$ are some dynamic characteristics of the neutral pion.
For the differential cross section it was obtained (in \cite{EK85} factor $1/16$ in the 
right-hand side of equation was lost):
\ba
d\sigma^{e\bar{e}\to \pi_0\pi_0\gamma}=\frac{1}{2!}\frac{\alpha}{16\pi^2m^4}G\frac{d^3q_1}{E_1}\frac{d^3q_2}{E_2}
\frac{d^3k}{\omega}\delta^4(p_++p_--q_1-q_2-k),
\ea
with
\ba
G=c_{11}|\alpha_1|^2+c_{22}|\alpha_2|^2+c_{33}|\alpha_3|^2+2c_{12}Re[\alpha_1\alpha_2^*]+2c_{13}Re[\alpha_1\alpha_3^*]+
2c_{32}Re[\alpha_3\alpha_2^*], \nn \\
\alpha_1=m^3a_1; \alpha_2=m^5a_2, \alpha_3=m^5a_3,
\ea
and
\ba
c_{11}=\frac{1}{E^2}(2\omega^2-\vec{k}^2); \nn \\
c_{22}=\frac{1}{E^2m^4}[8E^2Q_0^4+2\omega Q_0Q^2\vec{k}\vec{Q}-(\omega^2Q^2+4E^2Q_0^2)\vec{Q}^2-Q_0^2Q^2\vec{k}^2] \nn \\
c_{33}=\frac{4}{m^4}[8E^2Q_0^2-Q^2\vec{k}^2+4EQ_0\vec{k}\vec{Q}]; \nn \\
c_{12}=\frac{1}{E^2m^2}[-\omega^2\vec{Q}^2+2Q_0(\omega-E)\vec{k}\vec{Q}-q_0^2\vec{k}^2+4E\omega Q_0^2]; \nn \\
c_{13}=\frac{1}{E^2m^2}[8E^2\omega Q_0+2E\omega\vec{k}\vec{Q}-2EQ_0\vec{k}^2];\nn \\
c_{23}=\frac{1}{E^2M^4}[16E^3Q_0^3+2E(\omega Q^2+2EQ_0^2)\vec{k}\vec{Q}-4E^2Q_0\omega\vec{Q}^2-2EQ_0Q^2\vec{k}^2].
\ea
Here $\vec{q}_{1,2}, \vec{k}=-(\vec{q}_1+\vec{q}_2)$ are the two-dimensional vectors in the plane $x,y$
transversal to the beam axis $z$ direction
\ba
Q^2=m^2-\frac{1}{2}s(1-x), \,\,\,\,Q_0=\frac{1}{4}(x_1-x_2)\sqrt{s};\,\,\,\,\vec{k}^2=k_x^2+k_y^2; \nn \\
\vec{Q}^2=\frac{1}{4}[(q_{1x}-q_{2x})^2+(q_{1y}-q_{2y})^2];\,\,\,\,\vec{k}\vec{Q}=\frac{1}{2}[\vec{q}_2^2-\vec{q}_1^2].
\ea
Below, we will concentrate our attention on the Daliz
plot distribution over the energy fraction of pions.

The main mechanism of reaction includes two intermediate
vector meson states $V=\rho, \omega$:
\ba
\gamma^*\to \rho \to \omega \pi, \,\,\,\omega \to \gamma \pi^0,   \nn \\
\gamma^*\to \omega \to \rho \pi, \,\,\,\rho \to \gamma \pi^0.
\ea
The contribution of other possible channels is negligible \cite{PDG}
\ba
\Gamma_{\rho\to ee}\times B_{\omega\to \pi\gamma}=7 KeV\times 9\times 10^{-2}>>
\Gamma_{\omega\to ee}\times B_{\rho\to \pi\gamma}=0.8 KeV\times 6\times 10^{-4}.
\ea
We do not consider intermediate states with $\sigma(600)$ scalar meson as
well as an intermediate state with box-type quark loops. 
Almost complete cancelation of the relevant contributions, which takes place due 
to current algebra arguments commuting with the neutral current operators, 
was shown in \cite{box}.

We also remind that the emission of real photon by the initial leptons is strictly forbidden.
\section{Amplitude of the process of $e^- e^+ \to \gamma^*  \to \rho (\rho') \to \pi^0\omega \to \pi^0\pi^0\gamma$}
Using \cite{pigamm} we can derive the amplitude 
\ba
M=\frac{\sqrt{(4\pi\alpha)^3}g^2_{\rho\omega\pi}}{g_\rho^2}\frac{M_\rho^2}{[q^2-M_\rho^2+i\Gamma_\rho M_\rho]q^2}
J^\mu e(k)^\sigma T_{\mu\sigma}=M_\mu J^\mu,
\ea
where $g_\rho\approx 6$ is the $\rho$ meson decay constant, $M_\rho, \mbox{and} \Gamma_\rho$ are the mass 
and width of the $\rho$-meson and
\ba
T_{\mu\sigma}=\frac{V_{q1}V_{k2}}{s(1-x_1)-M_\omega^2+iM_\omega\Gamma_\omega}(\mu\lambda q q_1)(\lambda\sigma q_2 k)+
\frac{V_{q2}V_{k1}}{s(1-x_2)-M_\omega^2+iM_\omega\Gamma_\omega}(\mu\lambda q q_2)(\lambda\sigma q_1 k),
\ea
where we use the notation $(abcd)=\epsilon_{\alpha\beta\gamma\sigma}a^\alpha b^\beta c^\gamma d^\sigma$.
The quantities $V_{qi},V_{ki}$ take into account the deviation of the vertex $V^{\rho(q,\mu)\omega(p,\nu)\pi}_{\mu\nu}$
from its point-like approximation $g_{\rho\omega\pi}(\mu\nu q p)$;
$g_{\rho\omega\pi}\approx 16(GeV)^{-1}$ will be given in Appendix A.
\section{Dalitz-plot distribution}
\begin{figure}
\includegraphics[width=0.9\textwidth]{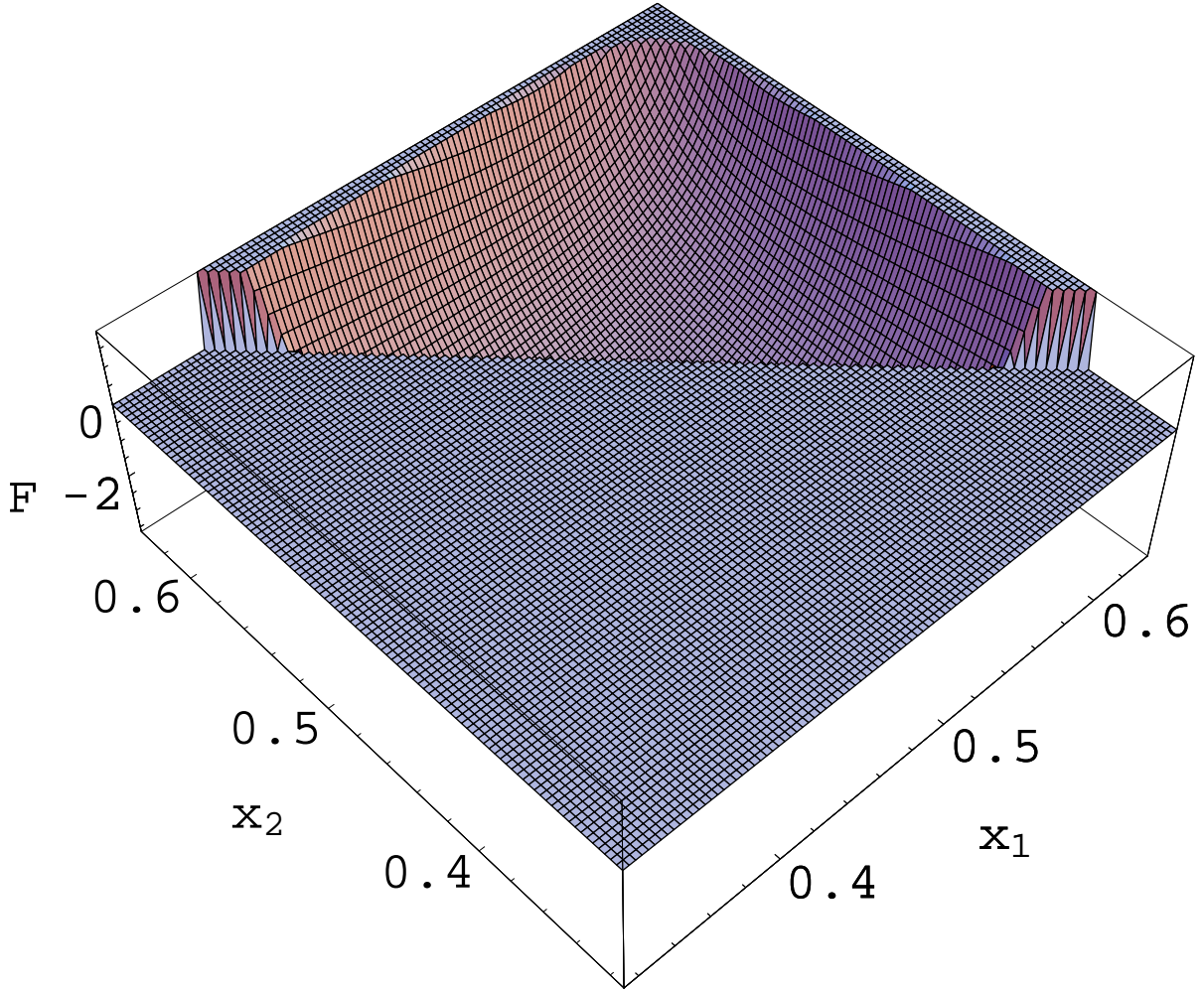}
\caption{Function $F(x_1, x_2)$ (see (16)) dependence of $x_1, x_2$.
\label{Fig1}}
\end{figure}

Applying the invariant integration method
\ba
\int M_\mu (M_\nu)^* d\Gamma=\frac{1}{3}(g_{\mu\nu}-\frac{q_\mu q_\nu}{q^2})\int |M_\lambda|^2 d\Gamma,
\ea
and using the expression for element of phase volume of the final state
\ba
d\Gamma=\frac{d^3 q_1}{2E_1}\frac{d^3 q_2}{2E_2}
\frac{d^3 k}{2\omega}\frac{1}{(2\pi)^5}\delta^4(p_++p_--q_1-q_2-k)=
\frac{s\pi^2}{(2\pi)^5}d x_1 d x_2\theta[(1-x_1)(1-x_2)(x_1+x_2-1)- \nn \\
(m^2/s)(4(1-x_1)(1-x_2)+(x_1-x_2)^2)],
\ea
we obtain for the Dalitz-plot distribution
\ba
\frac{d\sigma}{\sigma_0d x_1 d x_2}= F(x_1,x_2,s); \nn \\
F(x_1,x_2,s)=\frac{(1-x_1)^2(x-2(1-x_2))}{(1-x_1-M_\rho^2/s)^2}(V_{q1}V_{k2})^2+
\frac{(1-x_2)^2(x-2(1-x_1))}{(1-x_2-M_\rho^2/s)^2}(V_{q2}V_{k1})^2 +\nn \\
2\frac{(1-x_1)(1-x_2)(-1+x^2+x_1x_2)}{(1-x_1-M_\rho^2/s)(1-x_2-M_\rho^2/s)}
V_{q1}V_{k2}V_{q2}V_{k1}; \nn \\
\sigma_0=\frac{s^2g_{\rho\omega\pi}^4\alpha^3}{3(2\pi)^5g_\rho^4}.
\ea
The function $F(x_1,x_2)$ is presented in Fig.~\ref{Fig1}.
Since the set of tensors $L^{(i)}_{\mu\nu}$ is complete, we can present tensor $T_{\mu\nu}$ 
in the form
\ba
T=c_1L^{(1)}+c_2L^{(2)}+c_3L^{(3)},c_i=\frac{\Delta_i}{\Delta}, \nn \\
\Delta=\det|| L^{(i)}\times L^{(j)}||.
\ea
All the necessary convolutions needed to calculate $c_i$ are given in Appendix B.
The numerical calculation for $c_i(x_1,x_2)$ is presented in Tables II-IY, with $\rho$ meson
exchange in intermediate state. In Tables YI-YIII, we present the contribution of $\rho'$
meson to this quantities.

For completeness we put the Dalitz-plot distribution in terms of the bilinear combinations of $c_i$
\ba
\frac{d\sigma}{d x_1 d x_2}=\frac{1}{2!}\frac{4\alpha}{3s}[|\alpha_1|^2L_{11}+|\alpha_2|^2L_{22}+
|\alpha_3|^2L_{33}+2Re(\alpha_1\alpha_2^*L_{12}+\alpha_1\alpha_3^*L_{13}+\alpha_3\alpha_2^*L_{32})],
\ea
with the quantities $L_{ij}$ given in Appendix B.
\section{Total Cross Section}
For the energy region $300MeV<\sqrt{s}<1GeV$ the main contribution arises from intermediate states
with the $\rho,\omega$ mesons.
The total cross section
\ba
\sigma_{tot}(s)=\sigma_0(s)\int\limits_0^1 d x_2\int\limits_{1-x_2}^1 F(x_1,x_2,s) d x_1
\ea
is presented in Fig.4 where the experimental data \cite{Ahmet03} are presented as well.
Neglecting in this region the contributions of general polarizabilities related with
$a_2,a_3$ we can approximate the total cross section in the form
\ba
\sigma_{tot}(s)\approx \frac{\alpha}{12 M^2}(\alpha_{0}M^2\sqrt{s})^2 I(s), \nn \\
I(s)=\int\limits_0^{1-(4 m^2/s)} x^3\sqrt{1-\frac{4 m^2}{s(1-x)}} d x.
\ea
The electrical value of the neutral pion polarizability is estimated as
\ba
|\alpha_0|\approx 0,5\times 10^{-43}cm^3,
\ea
which is in reasonable agreement with that obtained in paper of one of us (MKV)
$\alpha_0\approx -065\times 10^{-43}cm^3$ \cite{MKV86}.

\begin{figure}
\includegraphics[width=0.9\textwidth]{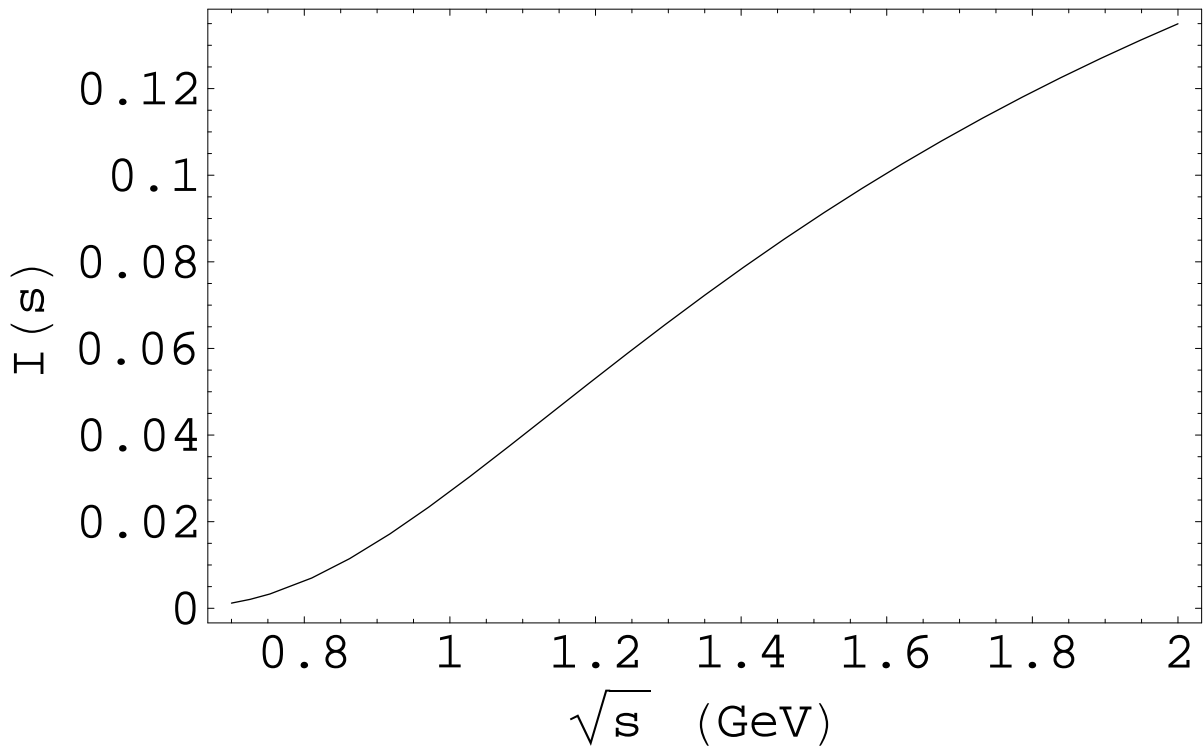}
\caption{Value $I(s)$ (defined in (16)) as a function of $s$.
\label{Fig2}}
\end{figure}

The function $I(s)$ is presented in Fig.~\ref{Fig2}.
The main contribution in the energy range $1GeV<\sqrt{s}<2GeV$ arises from a transition 
of the virtual photon to the ordinary $\rho$ and excited $\rho'$ mesons
\ba
M_\mu\sim \frac{1}{q^2}J^\mu[\frac{M_\rho^2}{q^2-M_\rho^2+iM_\rho\Gamma_\rho}+
C\frac{M_{\rho'}^2}{q^2-M_{\rho'}^2+iM_{\rho'}\Gamma_{\rho'}}],
\ea
with the $C$--relative phase factor.
\section{Disscusion and Conclusion}
The process  $e^-e^+ \to \pi^0\pi^0\gamma$ was experimentally studied, and the
total cross section was measured in \cite{Ahmet03}.

In this paper, we obtained the differential cross section and, in particular, the Dalitz-plot
distribution. The matrix element can be expressed in terms of generalized polarizabilities
of the neutral pion.  Some information on their values can be extracted from the distributions
obtained.

In Tables I-IY, the energy range $\sqrt {s} = 800 \,\,MeV$, where the $\rho$-meson exchange dominates.
In Tables Y-YIII, the energy was $\sqrt {s} 1.45 \,\,GeV$, where the $\rho'$- exchange dominates.
In both the cases the interference of the $\rho$ and $\rho'$ exchanges was shown to be negligible.

\begin{table}
\begin{tabular}{|c|c|c|c|c|c|c|c|c|c|c|c|c|}
\hline
$x_2 / x_1$ & 0.05 & 0.10 & 0.20 & 0.30 & 0.40 & 0.50 & 0.60 & 0.70 & 0.80 & 0.90 & 0.95  \\

\hline
0.95 & 2008.18 & 84.0845 & 12.247 & 5.04132 & 2.87478 & 1.92019 & 0.03241 & 0.02296 & 0.0031 &  &   \\
\hline
0.90 &  & 78.9729 & 11.3982 & 4.6616 & 2.6509 & 1.776 & 0.0111 & 0.0049 &  &  &   \\
\hline
0.80 &  &  & 9.4438 & 3.7205 & 2.0541 & 1.3605 &  &  &  &  & 0.00237  \\
\hline
0.70 &  &  &  & 2.538 & 1.2472 & 0.7529 &  &  &  & 0.00324 & 0.0142   \\
\hline
0.60 &  &  &  &  & 0.5216 & 0.1299 & 0 &  &  & 0.00529 & 0.0143  \\
\hline
0.50 &  &  &  &  &  & 2.033 & 0.0549 & 0.2277 & 0.3276 & 0.3579 & 0.3583   \\
\hline
0.40 &  &  &  &  &  &  & 0.3725 & 0.638 & 0.8365 & 0.9036 & 0.9074  \\
\hline
0.30 &  &  &  &  &  &  &  & 2.0646 & 2.4092 & 2.5266 & 2.5303  \\
\hline
0.20 &  &  &  &  &  &  &  &  & 9.3013 & 9.3964 & 9.3494   \\
\hline
0.10 &  &  &  &  &  &  &  &  &  & 95.7964 & 94.4592  \\
\hline
0.05 &  &  &  &  &  &  &  &  &  &  & 2709.61 \\
\hline
\end{tabular}
\caption{The function of $F(x_1,x_2)$ (see (16)), with $\rho$ meson exchange in 
intermediate state}
\label{Table1}
\end{table}
%
\begin{table}
\begin{tabular}{|c|c|c|c|c|c|c|c|c|c|c|c|c|}
\hline
$x_2 / x_1$ & 0.05 & 0.10 & 0.20 & 0.30 & 0.40 & 0.50 & 0.60 & 0.70 & 0.80 & 0.90 & 0.95  \\

\hline
0.95 & 0 & 0.0848 & 0.0932 & 0.10114 & 0.29932 & 0.3297 & -0.08686 & -0.1687 & -0.2145 & -0.2417 & -0.25034 \\
\hline
0.90 &  & 0 & 0.1397 & 4.8205 & 0.05259 & 0.2143 & -0.0621 & -0.1463 & -0.1941 & -0.2219 & -0.2417  \\
\hline
0.80 &  &  & 0 & 5.6177 & -0.03589 & 0.1204 & -0.02098 & -0.09602 & -0.14471 & -0.19409 & -0.21447  \\
\hline
0.70 &  &  &  & 0 & -0.12207 & 0.02772 & 0.02436 & -0.03866 & -0.09602 & -0.14626 & -0.16865  \\
\hline
0.60 &  &  &  &  & 0 & -0.10781 & 0.05765 & 0.02436 & -0.02098 & -0.0621 & -0.08686  \\
\hline
0.50 &  &  &  &  &  & 0 & -0.10781 & 0.02772 & 0.12039 & 0.2143 & 0.3297   \\
\hline
0.40 &  &  &  &  &  &  & 0 & -0.12207 & -0.03589 & 0.05258 & 0.2993   \\
\hline
0.30 &  &  &  &  &  &  &  & 0 & 5.6177 & 4.8205 & 0.10114   \\
\hline
0.20 &  &  &  &  &  &  &  &  & 0 & 0.13787 & 0.09317    \\
\hline
0.10 &  &  &  &  &  &  &  &  &  & 0 & 0.08481    \\
\hline
0.05 &  &  &  &  &  &  &  &  &  &  &  0 \\
\hline
\end{tabular}
\caption{The function of $C_1(x_1,x_2)$ (see (17))}
\label{Table1}
\end{table}
%
\begin{table}
\begin{tabular}{|c|c|c|c|c|c|c|c|c|c|c|c|c|}
\hline
$x_2 / x_1$ & 0.05 & 0.10 & 0.20 & 0.30 & 0.40 & 0.50 & 0.60 & 0.70 & 0.80 & 0.90 & 0.95  \\

\hline
0.95 & -23.5507 & -3.2455 & -0.05566 & 1.63001 & -3.49016 & 1.4138 & -0.8643 & -1.64083 & -1.9659 & -1.9406 & -1.8407 \\
\hline
0.90 &  & -5.7549 & -5.9489 & -246.831 & 6.8672 & 5.8997 & -1.4455 & -2.1446 & -2.1798 & -1.9809 & -1.9406  \\
\hline
0.80 &  &  & -6.9749 & -458.986 & 10.2559 & 8.3813 & -1.7504 & -2.3764 & -2.26102 & -2.1798 & -1.9659  \\
\hline
0.70 &  &  &  & -14.4509 & 16.8595 & 10.0958 & -1.8649 & -2.4164 & -2.3764 & -2.1464 & -1.6408  \\
\hline
0.60 &  &  &  &  & -68.7538 & 11.5783 & -1.8016 & -1.8649 & -1.7504 & -1.4455 & -0.8643  \\
\hline
0.50 &  &  &  &  &  & 0 & 11.5783 & 10.0958 & 8.3813 & 5.8997 & 1.4138   \\
\hline
0.40 &  &  &  &  &  &  & -68.7538 & 16.8595 & 10.2559 & 6.8672 & -3.49016   \\
\hline
0.30 &  &  &  &  &  &  &  & -14.4509 & -458.986 & -246.831 & 1.63001   \\
\hline
0.20 &  &  &  &  &  &  &  &  & -6.9749 & -5.9489 & -0.5566    \\
\hline
0.10 &  &  &  &  &  &  &  &  &  & -5.7549 & -3.2456    \\
\hline
0.05 &  &  &  &  &  &  &  &  &  &  & -23.5507 \\
\hline
\end{tabular}
\caption{The function of $C_2(x_1,x_2)$ (see (17))}
\label{Table1}
\end{table}
\begin{table}
\begin{tabular}{|c|c|c|c|c|c|c|c|c|c|c|c|c|c|}
\hline
$x_2 / x_1$ & 0.05 & 0.10 & 0.20 & 0.30 & 0.40 & 0.50 & 0.60 & 0.70 & 0.80 & 0.90 & 0.95  \\

\hline
0.95 & 0 & -0.55603 & -0.54972 & -0.54452 & -1.46007 & -1.49319 & 0.35126 & 0.63305 & 0.71704 & 0.5373 & 0 \\
\hline
0.90 &  & 0 & -0.93677 & -28.3278 & -0.34063 & -1.13183 & 0.24647 & 0.531003 & 0.5198 & 0 & -0.5373  \\
\hline
0.80 &  &  & 0 & -44.4922 & 0.02245 & -1.10212 & 0.03122 & 0.2426 & 0 & -0.5198 & -0.71704  \\
\hline
0.70 &  &  &  & 0 & 0.73466 & -1.1792 & -0.1842 & 0 & -0.2426 & -0.531 & -0.63305 \\
\hline
0.60 &  &  &  &  & 0 & -1.00014 & 0 & 0.1842 & -0.0312 & -0.2465 & -0.3513  \\
\hline
0.50 &  &  &  &  &  & 0 & 1.00014 & 1.17923 & 1.10212 & 1.13183 & 1.49319   \\
\hline
0.40 &  &  &  &  &  &  & 0 & -0.76346 & -0.02245 & 0.34063 & 1.46007  \\
\hline
0.30 &  &  &  &  &  &  &  & 0 & 44.4922 & 28.3278 & 0.54452   \\
\hline
0.20 &  &  &  &  &  &  &  &  & 0 & 0.93677 & 0.54972   \\
\hline
0.10 &  &  &  &  &  &  &  &  &  & 0 & 0.55603  \\
\hline
0.05 &  &  &  &  &  &  &  &  &  &  & 0 \\
\hline
\end{tabular}
\caption{The function of $C_3(x_1,x_2)$ (see (17))}
\label{Table1}
\end{table}

\newpage

\begin{table}
\begin{tabular}{|c|c|c|c|c|c|c|c|c|c|c|c|c|}
\hline
$x_2 / x_1$ & 0.05 & 0.10 & 0.20 & 0.30 & 0.40 & 0.50 & 0.60 & 0.70 & 0.80 & 0.90 & 0.95  \\

\hline
0.95 & 8.548$\cdot 10^{-6}$ & 7.98$\cdot 10^{-5}$ & 8.522$\cdot 10^{-4}$ & 0.0018 & 0.00269 & 0.00322 & 0.00375 & 0.00373 & 0.0035 &  &   \\
\hline
0.90 &  & 7.57$\cdot 10^{-5}$ & 7.95$\cdot 10^{-4}$ & 1.646$\cdot 10^{-3}$ & 2.39$\cdot 10^{-3}$ & 2.72$\cdot 10^{-3}$ & 2.88$\cdot 10^{-3}$ & 2.29$\cdot 10^{-3}$ & 7.21$\cdot 10^{-4}$ &  &   \\
\hline
0.80 &  &  & 0.000734 & 0.00139 & 0.00174 & 0.00137 & 8.8$\cdot 10^{-5}$ &  &  & 5.25$\cdot 10^{-4}$ & 0.00176  \\
\hline
0.70 &  &  &  & 0.00139 & 0.0016 & 0.000099 &  &  &  & 0.00636 & 0.0071   \\
\hline
0.60 &  &  &  &  & 0.002 & 0.0015 & 0 &  & 0.00097 & 0.023 & 0.021  \\
\hline
0.50 &  &  &  &  &  & 0.0036 & 0.0042 & 0.0079 & 0.0419 & 0.0605 & 0.0493   \\
\hline
0.40 &  &  &  &  &  &  & 0.0149 & 0.0345 & 0.1419 & 0.1418 & 0.1097  \\
\hline
0.30 &  &  &  &  &  &  &  & 0.088 & 0.339 & 0.2915 & 0.219  \\
\hline
0.20 &  &  &  &  &  &  &  &  & 0.7533 & 0.5946 & 0.4374   \\
\hline
0.10 &  &  &  &  &  &  &  &  &  & 1.0366 & 0.7503  \\
\hline
0.05 &  &  &  &  &  &  &  &  &  &  & 1.0688 \\
\hline
\end{tabular}
\caption{The function of $F(x_1,x_2)$ (see (16)), with $\rho'$ meson exchange in
intermediate state}
\label{Table1}
\end{table}
%
\begin{table}
\begin{tabular}{|c|c|c|c|c|c|c|c|c|c|c|c|c|}
\hline
$x_2 / x_1$ & 0.05 & 0.10 & 0.20 & 0.30 & 0.40 & 0.50 & 0.60 & 0.70 & 0.80 & 0.90 & 0.95  \\

\hline
0.95 & 0 & 0.000949 & 0.0052 & 0.0127 & 0.0199 & 0.046 & 0.0983 & 0.212 & 0.534 & 0.1897 & 0.0825 \\
\hline
0.90 &  & 0 & 0.01176 & 1.0559 & -0.0188 & 0.0142 & 0.0684 & 0.1816 & 0.4565 & 0.1916 & 0.1897  \\
\hline
0.80 &  &  & 0 & 3.095 & -0.119 & -0.083 & -0.0216 & 0.1154 & 0.358 & 0.456 & 0.534  \\
\hline
0.70 &  &  &  & 0 & -0.1067 & -0.0808 & -0.032 & 0.0471 & 0.1154 & 0.182 & 0.212  \\
\hline
0.60 &  &  &  &  & 0 & -0.09114 & -0.06059 & -0.032 & -0.0216 & 0.0684 & 0.0983  \\
\hline
0.50 &  &  &  &  &  & 0 & -0.09114 & -0.0808 & -0.0831 & 0.0142 & 0.0461   \\
\hline
0.40 &  &  &  &  &  &  & 0 & -0.1067 & -0.1192 & -0.0188 & 0.01995   \\
\hline
0.30 &  &  &  &  &  &  &  & 0 & 3.0951 & 1.0559 & 0.0127   \\
\hline
0.20 &  &  &  &  &  &  &  &  & 0 & 0.1175 & 0.0052    \\
\hline
0.10 &  &  &  &  &  &  &  &  &  & 0 & 0.00095    \\
\hline
0.05 &  &  &  &  &  &  &  &  &  &  &  0 \\
\hline
\end{tabular}
\caption{The function of $C_1(x_1,x_2)$ (see (17))}
\label{Table1}
\end{table}
%
\begin{table}
\begin{tabular}{|c|c|c|c|c|c|c|c|c|c|c|c|c|}
\hline
$x_2 / x_1$ & 0.05 & 0.10 & 0.20 & 0.30 & 0.40 & 0.50 & 0.60 & 0.70 & 0.80 & 0.90 & 0.95  \\

\hline
0.95 & -0.0174 & -0.0394 & -0.0488 & -0.0496 & 0.0576 & 0.0994 & 0.263 & 0.6044 & 1.458 & 0.4615 & 0.1846 \\
\hline
0.90 &  & -0.164 & -0.3466 & -16.87 & 0.417 & 0.3257 & 0.4613 & 0.7912 & 1.528 & 0.5207 & 0.4615  \\
\hline
0.80 &  &  & -1.6412 & -78.913 & 0.9049 & 0.4274 & 0.5603 & 1.0157 & 1.7026 & 1.5277 & 1.4579  \\
\hline
0.70 &  &  &  & -2.887 & 1.353 & 0.4071 & 0.6085 & 0.8952 & 1.0157 & 0.7912 & 0.6044  \\
\hline
0.60 &  &  &  &  & -10.3302 & 0.3966 & 0.5765 & 0.6085 & 0.5603 & 0.4613 & 0.2633  \\
\hline
0.50 &  &  &  &  &  & 0 & 0.3967 & 0.4071 & 0.4274 & 0.3257 & 0.099   \\
\hline
0.40 &  &  &  &  &  &  & -10.3302 & 1.3531 & 0.9049 & 0.4171 & 0.05758   \\
\hline
0.30 &  &  &  &  &  &  &  & -2.887 & -78.9132 & -16.8713 & -0.0496   \\
\hline
0.20 &  &  &  &  &  &  &  &  & -1.6412 & -0.3466 & -0.04876    \\
\hline
0.10 &  &  &  &  &  &  &  &  &  & -0.164 & -0.0349    \\
\hline
0.05 &  &  &  &  &  &  &  &  &  &  & -0.01739 \\
\hline
\end{tabular}
\caption{The function of $C_2(x_1,x_2)$ (see (17))}
\label{Table1}
\end{table}
\begin{table}
\begin{tabular}{|c|c|c|c|c|c|c|c|c|c|c|c|c|c|}
\hline
$x_2 / x_1$ & 0.05 & 0.10 & 0.20 & 0.30 & 0.40 & 0.50 & 0.60 & 0.70 & 0.80 & 0.90 & 0.95  \\

\hline
0.95 & 0 & -0.00188 & -0.00924 & -0.02039 & -0.02943 & -0.0627 & -0.1242 & -0.2505 & -0.5945 & -0.1737 & 0 \\
\hline
0.90 &  & 0 & -0.0237 & -1.886 & 0.0314 & -0.0179 & -0.086 & -0.2135 & -0.507 & 0 & 0.1737  \\
\hline
0.80 &  &  & 0 & -7.445 & 0.2545 & 0.1893 & 0.1388 & 0.0713 & 0 & 0.5073 & 0.5945  \\
\hline
0.70 &  &  &  & 0 & 0.297 & 0.179 & 0.08793 & 0 & -0.0712 & 0.21348 & 0.2505 \\
\hline
0.60 &  &  &  &  & 0 & 0.1885 & 0 & -0.0879 & -0.1387 & 0.086 & 0.124  \\
\hline
0.50 &  &  &  &  &  & 0 & -0.1885 & -0.179 & -0.189 & 0.0179 & 0.0627   \\
\hline
0.40 &  &  &  &  &  &  & 0 & -0.297 & -0.2545 & -0.03136 & 0.0294  \\
\hline
0.30 &  &  &  &  &  &  &  & 0 & 7.4449 & 1.8859 & 0.0204   \\
\hline
0.20 &  &  &  &  &  &  &  &  & 0 & 0.0237 & 0.00924   \\
\hline
0.10 &  &  &  &  &  &  &  &  &  & 0 & 0.00188  \\
\hline
0.05 &  &  &  &  &  &  &  &  &  &  & 0 \\
\hline
\end{tabular}
\caption{The function of $C_3(x_1,x_2)$ (see (17))}
\label{Table1}
\end{table}
\section{Appendix A}

Deviation from the point approximation for the coupling constant $g_{\rho\omega\pi}$ is due to the
three-angle loop Feynman integral
\ba
I_{\mu\nu}(q,q_1)=\frac{4\pi^2N_c}{(2\pi)^4}\int\frac{d^4 k}{i\pi^2}\frac{Sp}{(k^2-M^2)((k+q_1)^2-M^2)
((k+q_1-q)^2-M^2)}, \nn \\
Sp=\frac{1}{4}Sp(\hat{k}+M)\gamma_\nu(\hat{k}+\hat{q}_1+M)\gamma_\mu(\hat{k}+\hat{q}_1-\hat{q}+M)\gamma_5,
\ea
with $M$-constituent quark mass $M\approx 280MeV$.
The standard procedure of joining the denominators performing the loop momentum integration leads to
\ba
I_{\mu\nu}(q,q_1)=g_{\rho\omega\pi}(\mu\nu q (q-q_1))/g_\rho^2 V_{q,1},g_\rho\approx 6,
\ea
with
\ba
g_{\rho\omega\pi}=\frac{g_\rho^2N_c}{8\pi^2M}\approx \frac{16}{GeV},
\ea
and
\ba
V_{q,1}=2M^2\int\frac{d^3 x\delta(x_1+x_2+x_3-1)}{M^2-q^2x_2x_3-q_1^2x_1x_2-(q-q_1)^2x_1x_3}.
\ea
Note that in the limit of a heavy quark mass we obtain $V_{q,1}=1$.
In the approximation of zero pion mass we have for quantities used above
\ba
V_{q1}=\frac{2M^2}{sx_1}\int\limits_0^1\frac{dz}{z}\ln|\frac{1-z(1-z)s(1-x_1)/M^2}{1-z(1-z)s/M^2}|; \nn \\
V_{k2}=-\frac{2M^2}{s(1-x_1)}\int\limits_0^1\frac{d z}{z}\ln|1-z(1-z)s(1-x_1)/M^2|.
\ea
The remaining quantities $V_{q2},V_{k1}$ can be obtained from these expressions by the replacement
$x_1\to x_2$.
To provide the QCD principle of quark confinement, we neglect the possible imaginary parts of the
three-angle amplitude.

\section{Appendix B}

The convolutions of the tensors $L^{(i)}\times L^{(j)}$ are (indices suppressed):
\ba
L_{11}=\frac{1}{s^2}L^{(1)}\times L^{(1)}=\frac{1}{2}x^2; \nn \\
L_{22}=\frac{1}{s^4}L^{(2)}\times L^{(2)}=\frac{1}{128}[8\bar{x}^2\bar{x}_1\bar{x}_2+(x_1-x_2)^2]; \nn \\
L_{33}=\frac{1}{s^4}L^{(3)}\times L^{(3)}=\frac{1}{16}[(1+x)(x_1-x_2)^2+x^2\bar{x}]; \nn \\
L_{12}=\frac{1}{s^3}L^{(1)}\times L^{(2)}=\frac{1}{16}[-x^2\bar{x}+(x_1-x_2)^2]; \nn \\
L_{13}=\frac{1}{s^3}L^{(1)}\times L^{(3)}=\frac{1}{4}x(x_1-x_2); \nn \\
L_{23}=\frac{1}{s^4}L^{(2)}\times L^{(3)}=\frac{1}{32}(x_1-x_2)^3.
\ea
The conversion of the tensors $L^{(i)}$ with the tensor $T$ are
\ba
L^{(1)}\times T=\frac{s^2}{4}g_{\rho\omega\pi}[\frac{x\bar{x}-\bar{x}_2}{1-(M_\omega^2/s)}
V_{q1}V_{k2}+(x_1 \leftrightarrow x_2)]; \nn \\
L^{(2)}\times T=\frac{s^3}{32}g_{\rho\omega\pi}[\frac{x\bar{x}\bar{x}_2-(x-x_2)(x_1-x_2)^2}
{1-(M_\omega^2/s)}V_{q1}V_{k2}+(x_1 \leftrightarrow x_2)]; \nn \\
L^{(3)}\times T=\frac{s^3}{8}g_{\rho\omega\pi}[\frac{(x-x_2)(x_1-x_2)-\bar{x}_2\bar{x}}
{1-(M_\omega^2/s)}V_{q1}V_{k2}-(x_1 \leftrightarrow x_2)].
\ea
\acknowledgements

We are grateful to A.B.Arbuzov for useful discussions.
This work was supported by RFBR, grant no. 10-02-01295a.
This work was also supported by the Heisenberg--Landau program grant HLP-2010-06, and
JINR--Belorus--2010 grant.

\end{document}